\documentclass[10pt, conference]{IEEEtran}

\usepackage{amsmath,amssymb,amsfonts}
\usepackage{pgfplots}
\pgfplotsset{compat=1.18}
\usepackage{graphicx}
\usepackage{textcomp}
\usepackage{xcolor}
\usepackage{mathtools}

\usepackage{algorithm}
\usepackage[noEnd=true]{algpseudocodex}
\newcommand{\To}{\textbf{to }}

\usepackage{booktabs}
\usepackage{fancyvrb} 
\usepackage{multirow}
\usepackage{diagbox}
\usepackage{tikz}
\usepackage{numprint}
\usepackage{url}
\usepackage{colortbl}
\usepackage{dhucs} 
\usepackage{subcaption}
\DeclareCaptionFont{customsize}{\fontsize{9.5pt}{10.5pt}\selectfont}
\usepackage{enumitem} 
\usepackage{makecell} 
\usepackage[noadjust]{cite} 

\newcommand\refFigure[1]{Fig.~\ref{#1}}
\newcommand\refTable[1]{Table~\ref{#1}}

\newcommand\korean[1]{}
\newcommand\blind[1]{XXXX}

\renewcommand{\baselinestretch}{0.97}
\IEEEoverridecommandlockouts
\IEEEaftertitletext{\vspace{-0.1\baselineskip}}     

\begin{document}
\bstctlcite{BSTcontrol}

\title{MASQ: Measurement-Adaptive Sampling for Near-Clifford Quantum Error Correction Circuits}
\IEEEaftertitletext{\vspace{-3ex}}

\author{
Chaebin Jung, Sangmin Jeon, Kyeongwon Lee, Sechan Park, and Woojoo Lee\\
School of Intelligent Semiconductor Engineering, Chung-Ang University, Seoul, Korea \\

\thanks{
This work was supported in part by the National Research Foundation of Korea (NRF) grant funded by the Korea government (MSIT) (No. RS-2024-00345668). Corresponding author: Woojoo Lee

Chaebin Jung and Sangmin Jeon Lee contributed equally to this work.


}
}
\maketitle

\begin{abstract}
\boldmath
Classical sampling of quantum error correction (QEC) circuits supports fault-tolerant protocol evaluation, but estimating rare logical failures requires many independent shots.
Frame--active-state simulators isolate non-stabilizer information in a dense active-state vector, yet dense-state processing remains exponential in the active width.
This paper presents \textit{MASQ}, an exact measurement-adaptive sampler for near-Clifford QEC circuits based on selective materialization of non-Clifford rotations.
At each measurement boundary, MASQ materializes only rotations in a measurement-relevant closure capturing direct and indirect non-commuting dependencies, deferring the rest to limit active-state expansion.
Stabilizer-assisted elimination further removes rotations certified to be invisible to all future measurements.
A circuit-dependent schedule is computed once and shared across shots, while adaptive caching of conditional Born probabilities and batched execution reduce repeated sampling work while preserving the measurement-record distribution.
Across 19 QEC benchmarks, MASQ reduces peak active width by up to 40 active virtual qubits and achieves up to $8{,}098\times$ CPU throughput speedup over Clifft on multi-round circuits.
\unboldmath
\end{abstract}

\section{Introduction}
Quantum error correction (QEC) protects logical information by encoding it across physical qubits, enabling fault-tolerant quantum computing (FTQC)~\cite{Schmid:DATE2025, Stade:DATE2025, Lao:PRA2020}.
Because large-scale, repeated hardware evaluation remains challenging, classical simulation is widely used to assess protocol reliability~\cite{Zhang:DATE2026, Wang:ISCA2025, Google:Nature2025, Bausch:Nature2024}.
Direct calculation of measurement probabilities or logical error rates (LERs) under a noise model can be costly as fault and outcome combinations grow with circuit size~\cite{Cicero:TQC2025, Xu:Bull2025, Chen:CAV2025}.
Sampling instead tracks one realization of noise and measurement outcomes per \emph{shot}, using repeated shots to estimate LER statistically~\cite{Gidney:Quantum2021, Chase:arXiv2026, Fang:arXiv2026}.
In the fault-tolerant regime, rare logical failures require many shots~\cite{Heussen:PRR2024, Google:Nature2025, Paler:Quantum2023, GarciaHerrero:Quantum2026}, making shot-generation throughput critical.

Clifford circuits with Pauli noise and measurements admit compact stabilizer-tableau representations~\cite{Aaronson:PRA2004}.
Stim~\cite{Gidney:Quantum2021} exploits this structure by preprocessing deterministic Clifford structure and propagating stochastic Pauli errors and measurement-result changes.
However, FTQC requires non-Clifford resources such as $T$ gates, which arise in magic-state injection, distillation, and cultivation.
These resources and coherent noise can generate non-stabilizer information that a single tableau or Pauli frame cannot represent, making exact near-Clifford sampling substantially more expensive~\cite{Chen:CAV2025}.

Clifford-compatible or stochastic Pauli approximations retain efficient stabilizer sampling~\cite{Singh:NPJ2026, Bausch:Nature2024} but can bias LER and threshold estimates by altering non-Clifford dynamics~\cite{Garner:PRR2026, Schwartzman:PRA2025}.
Exact methods avoid this approximation error but remain limited by representation costs~\cite{Mocherla:arXiv2023}.
State-vector methods~\cite{Katsuda:PRR2024, Itogawa:PRXQ2025} retain general amplitudes but require $O(2^N)$ memory for $N$ qubits~\cite{Choi:TC26, Park:ICCAD26, Choi:QIP26}.
Tensor-network costs~\cite{Orioli:arXiv2026, Manabe:NJP2025} depend on peak intermediate-tensor width, while finding an optimal contraction order is NP-hard~\cite{Berezutskii:NRP2025, Burgholzer:TCAD2023}.
Stabilizer-derived methods~\cite{Haenel:arXiv2026, Tuloup:arXiv2026} avoid the full state vector, but their costs grow with stabilizer rank or the number of nonzero generalized-stabilizer coefficients~\cite{Li:arXiv2025, Tuloup:arXiv2026}.

Frame--active-state hybrid methods isolate non-stabilizer degrees of freedom in a dense active vector while handling Clifford--Pauli structure through compact frames.
Clifft~\cite{Chase:arXiv2026} maintains a frame-factored state with $2^k$ amplitudes for $k$ active virtual qubits, and SymFT~\cite{Fang:arXiv2026} reduces per-shot frame updates and bookkeeping through symbolic Clifford--Pauli factorization and offline stabilizer-coordinate planning.
Nevertheless, dense-state processing remains $O(2^k)$, making peak active width $k_{\max}$ a central bottleneck.

Our key observation is that not all unresolved non-Clifford information must be materialized to evaluate the current measurement.
We propose \textbf{MASQ} (Measurement-Adaptive Sampling for QEC), an exact sampler for near-Clifford QEC circuits.
MASQ keeps rotations pending, constructs a \emph{measurement-relevant closure} at each boundary, and materializes only the rotations in that closure.
The closure captures both direct measurement dependencies and earlier rotations connected through non-commuting dependencies, while other rotations remain pending for later measurements.
MASQ further uses stabilizer-equivalent relations to eliminate \emph{measurement-invisible} rotations whose effects can be moved beyond all future measurements.
Together, they reduce avoidable active-state growth and exponential cost.

By retaining only measurement-relevant non-Clifford information at each boundary, MASQ exposes recurring probability-determining contexts that adaptive caching exploits to reuse conditional Born probabilities.
These optimizations preserve conditional Born probabilities at every boundary and therefore the joint measurement-record distribution.
Across 19 QEC benchmarks, MASQ reduces peak active width by up to 40 active virtual qubits and achieves CPU throughput gains up to $8{,}098\times$ over Clifft and $1{,}267\times$ over CPU SymFT on multi-round surface-code circuits, while completing circuits where existing exact simulators hit memory or compilation limits.
A numerical correctness validation further confirms that MASQ reproduces baseline Born probabilities within {$1.17{\times}10^{-17}$} across 15 benchmarks with available baselines.

\vspace{-0.2em}
\section{Background and Motivation}\label{sec:preliminaries}
\vspace{-0.2em}
\subsection{Frame-Factored State Representation}\label{sec:frame}

Frame-factored simulation separates Clifford--Pauli transformations from quantum information that must be tracked explicitly using dense amplitudes.
At any point along a shot trajectory, an $N$-qubit physical state $|\Psi\rangle$ is represented as
\begin{equation}
|\Psi\rangle = U_C P_F \left(|\phi\rangle_A \otimes |0\rangle_D\right),
\label{eq:frame-state}
\end{equation}
where $U_C$ is the Clifford frame representing deterministic Clifford operations and $P_F$ is the Pauli frame containing shot-dependent Pauli errors and corrections.
The virtual qubits are partitioned into an active set $A$ and a dormant set $D$, where $A\cap D=\emptyset$ and $|A|+|D|=N$.
The active state $|\phi\rangle_A$ stores the remaining quantum information, whereas each dormant coordinate is fixed to $|0\rangle$ in the virtual basis.

Each dormant coordinate $v$ satisfies $Z_v|0\rangle_v=|0\rangle_v$.
More generally, if a state $|\chi\rangle$ satisfies $S|\chi\rangle=\epsilon|\chi\rangle$ for a Hermitian Pauli operator $S$ and $\epsilon\in\{+1,-1\}$, then $\epsilon S$ is a stabilizer of $|\chi\rangle$~\cite{Gidney:Quantum2021}.
The Pauli stabilizers of a state form a commuting group with independent generators, while information not captured by these relations is retained in the dense active state.
For an active width $k=|A|$, $|\phi\rangle_A$ contains $2^k$ complex amplitudes rather than the $2^N$ amplitudes of a full state vector.

Because the active state is expressed in the virtual basis defined by $U_C$, physical Pauli axes must first be mapped into this basis.
For physical rotation and measurement axes $Q_{\rm phys}$ and $M_{\rm phys}$, respectively, the corresponding virtual axes are
\begin{equation}
Q=U_C^\dagger Q_{\rm phys}U_C, \qquad M=U_C^\dagger M_{\rm phys}U_C.
\label{eq:heisenberg}
\end{equation}
Since $U_C$ is Clifford, the mapped axes remain signed Pauli operators~\cite{Gottesman:arXiv1998}.
A deterministic Clifford gate $C$ is absorbed into the Clifford frame as $U_C\leftarrow CU_C$, whereas a physical Pauli operation $E$ is mapped to the virtual basis and accumulated as $P_F\leftarrow(U_C^\dagger E U_C)P_F$.
Neither update directly modifies $|\phi\rangle_A$ or changes the active width $k$.

A nonidentity virtual Pauli string can be mapped to a single virtual coordinate through \emph{Pauli localization}~\cite{Li:ASPLOS2022}:
\begin{equation}
VQV^\dagger=\eta\sigma_v,~~\sigma_v\in\{X_v,Y_v,Z_v\},~~\eta\in\{-1,+1\}.
\label{eq:localization}
\end{equation}
The same localization applies to a measurement axis $M$.
Here, $V$ is a change of virtual coordinates rather than an additional physical gate, and the corresponding representation is updated consistently so that the physical state in Eq.~\eqref{eq:frame-state} remains unchanged.
Below, $Q$ and $M$ denote the signed Pauli axes in the current virtual coordinates.

\subsection{Non-Clifford Rotations and Pauli Measurements}\label{sec:active-update}

Unlike Clifford and Pauli operations, a non-Clifford rotation can introduce non-stabilizer information that cannot be absorbed into either frame.
A rotation about a Hermitian Pauli axis $Q$ is defined as
\begin{equation}
R_Q(\theta)=e^{-i\theta Q/2}.
\label{eq:pauli-rotation}
\end{equation}
Its interaction with the Pauli frame follows
\begin{equation}
R_Q(\theta)P_F
=
P_F R_Q\left((-1)^{\omega(Q,P_F)}\theta\right),
\label{eq:rotation-frame}
\end{equation}
where $\omega(\cdot,\cdot)\in\{0,1\}$ is the commutation parity, with $0$ and $1$ denoting commutation and anticommutation, respectively.

If the localized coordinate $v$ is already active, the rotation updates the existing $2^k$ amplitudes without changing $k$.
If $v$ is dormant and the rotation introduces non-stabilizer information, the simulator promotes it to the active set as $A\leftarrow A\cup\{v\}$, $D\leftarrow D\setminus\{v\}$, and $k\leftarrow k+1$, doubling the dense active-state size from $2^k$ to $2^{k+1}$.

Pauli measurements can conversely reduce the active width.
Let $\tilde b\in\{0,1\}$ denote the measurement outcome in the virtual-state representation, with projector $\Pi_{\tilde b}(M)=[I+(-1)^{\tilde b}M]/2$ for a virtual measurement axis $M$.
When the localized measurement acts on an active coordinate, its Born probability is $p_{\tilde b}=\|\Pi_{\tilde b}(M)|\phi\rangle_A\|^2$, where identity factors outside $A$ are omitted.
The simulator samples $\tilde b$ from $\{p_0,p_1\}$ and updates the active state as $|\phi\rangle_A\leftarrow\Pi_{\tilde b}(M)|\phi\rangle_A/\sqrt{p_{\tilde b}}$.
The recorded physical outcome is $b=\tilde b\oplus\omega(M,P_F)$, where $P_F$ is evaluated immediately before measurement, so an anticommuting Pauli frame reverses the sampled virtual outcome.
  
After an active coordinate is measured, it can be returned to the dormant set once the post-measurement Pauli eigenstate is factored out and the frame representation is updated to restore the dormant-state convention $|0\rangle_v$.
The simulator then moves $v$ from $A$ to $D$ and decreases $k$ by one, halving the dense active-state size from $2^k$ to $2^{k-1}$.
    
If a localized measurement instead acts on a dormant coordinate, the outcome is determined or sampled from its stabilizer relation and the resulting state change can be represented through Clifford--Pauli frame updates without dense-state projection or an increase in $k$.
Consequently, Clifford and Pauli operations leave the active width unchanged, whereas non-Clifford rotations and measurements can introduce and remove active degrees of freedom, respectively.

\begin{figure*}[t]
\vskip -6pt
\centering
\includegraphics[width=\textwidth]{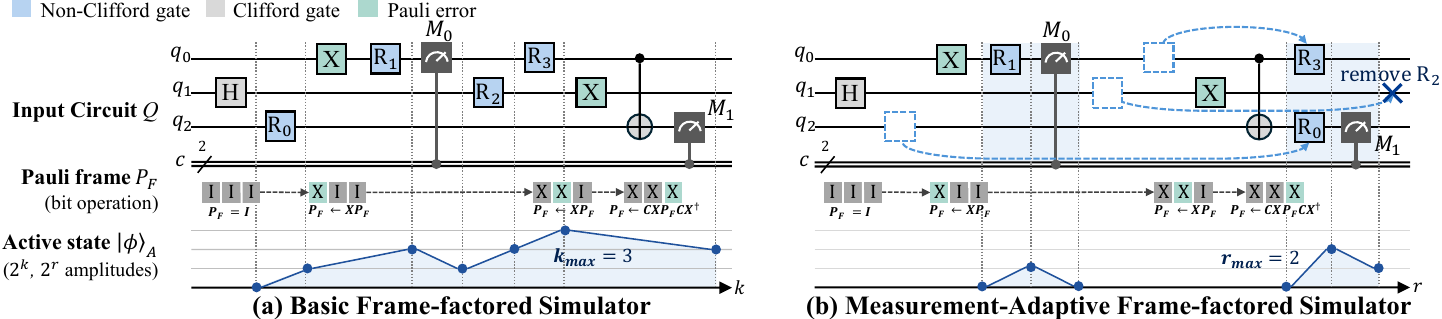}
\caption{Illustrative materialization schedules for frame-factored simulation.
(a) An eager execution without commutation-based scheduling.
(b) MASQ materializes only rotations in each measurement-relevant closure and eliminates those certified to be invisible to future measurements, reducing the peak active width from $k_{\max}=3$ to $r_{\max}=2$ in this example.}
\label{fig:overview}
\vskip -8pt
\end{figure*}

\vspace{-0.2em}
\subsection{Active-Width Bottleneck and Measurement Relevance}\label{sec:active-bottleneck}
\vspace{-0.2em}

The exponential component of frame-factored simulation is governed by the active width $k$ rather than directly by the total physical-qubit count $N$.
Let $k_{\max}$ denote the maximum active width during circuit execution.
The dense active state requires $O(2^{k_{\max}})$ peak memory, while a non-Clifford rotation, Born probability evaluation, or projection on a width-$k$ active state requires $O(2^k)$ dense-state work.
Total execution time additionally depends on the number of such operations and on frame-management and localization overheads.

When $k_{\max}\ll N$, frame-factored simulation confines the exponential state representation to a much smaller subspace than the full $2^N$ state vector.
Its cost nevertheless remains exponentially sensitive to active-state growth, particularly when non-Clifford rotations accumulate before measurements remove their associated active degrees of freedom.
For example, Clifft reduces $k_{\max}$ through equivalence-preserving Heisenberg IR (HIR) reordering that pushes commuting non-Clifford operations later and measurements earlier~\cite{Chase:arXiv2026}.
Even with this optimization, its coherent-noise QEC evaluation reports $k_{\max}$ values from $5$ to $24$, over which sampling throughput decreases by several orders of magnitude.
These results highlight that active-state growth remains a central scalability bottleneck.

This motivates reducing active-state work beyond commutation-based scheduling when only exact measurement statistics, rather than the full quantum state, must be preserved~\cite{Paykin:QCE2023, Schmitz:arXiv2023}.
MASQ combines measurement-relevant closures at measurement boundaries with stabilizer-assisted elimination, which uses state-dependent equivalent axes to certify rotations as invisible to all future measurements.
The resulting probability-determining contexts enable adaptive reuse of conditional Born probabilities across noisy shots.

\section{MASQ: Proposed Method}\label{sec:proposed}
Fig.~\ref{fig:overview} illustrates the materialization behavior of MASQ relative to an eager frame-factored execution without commutation-based scheduling.
In the eager execution, non-Clifford rotations are materialized as encountered, whereas MASQ keeps rotations pending and materializes only those required at each measurement boundary in their original circuit order.
Rotations irrelevant to the current measurement remain pending, while those certified to be invisible to all future measurements are eliminated.
The following analysis establishes when these transformations preserve the conditional Born probabilities and the measurement-record distribution.

\subsection{Measurement-Adaptive Active-State Reduction}\label{sec:masq_prove}
Let $m$ denote the number of measurements, $M_n$ the virtual Pauli axis of the $n$th measurement, and $G_n$ the gate segment between $M_{n-1}$ and $M_n$, with $G_1$ starting at circuit initialization.
As defined in Section~\ref{sec:active-update}, $\tilde b_n$ denotes the virtual outcome used for active-state projection, whereas $b_n$ denotes the recorded physical measurement bit.
A shot produces the measurement record $\mathbf b=(b_1,\ldots,b_m)$ by sequentially sampling conditional Born probabilities.
All commutation tests below are performed with the corresponding rotation and measurement axes expressed in the same virtual coordinates.

\noindent\textbf{Commutation-Based Rotation Deferral.}
Consider a pending rotation $R_Q(\vartheta)$, where $\vartheta$ denotes its effective angle including the shot-dependent Pauli-frame sign.
If $\omega(Q,M_n)=0$, the rotation commutes with either measurement projector:
\begin{equation}
\Pi_{\tilde b}(M_n)R_Q(\vartheta)=R_Q(\vartheta)\Pi_{\tilde b}(M_n).
\label{eq:commute}
\end{equation}
For a normalized virtual state $|\psi\rangle$, the Born probability of outcome $\tilde b$ is therefore
\begin{equation}
\Pr(\tilde b)=\langle\psi|R_Q^\dagger(\vartheta)\Pi_{\tilde b}(M_n)R_Q(\vartheta)|\psi\rangle
=\langle\psi|\Pi_{\tilde b}(M_n)|\psi\rangle.
\label{eq:commute-born}
\end{equation}
Thus, when considered in isolation, a rotation commuting with $M_n$ can be moved beyond the measurement without changing its conditional Born probabilities~\cite{Paykin:QCE2023}, provided that its effect remains pending afterward.
Pauli-frame propagation can change the effective angle sign according to Eq.~\eqref{eq:rotation-frame}, but does not change the Pauli commutation relations used for dependency analysis.

\noindent\textbf{Measurement-Relevant Closure.}
Commutation with $M_n$ alone is insufficient when multiple rotations are pending because an earlier rotation can affect the measurement indirectly through a later non-commuting rotation.
Let $\mathcal L_n$ denote the ordered set of rotation indices pending immediately before $M_n$, where the indices preserve their original circuit order.
MASQ first identifies the rotations that directly anticommute with the measurement:
\begin{equation}
\mathcal C_n^{(0)}\coloneqq\left\{a\in\mathcal L_n\mid\omega(Q_a,M_n)=1\right\}.
\label{eq:closure-seed}
\end{equation}
It then recursively adds an earlier rotation whenever it anticommutes with an already selected later rotation:
\begin{equation}
\mathcal C_n^{(s+1)}=\mathcal C_n^{(s)}
\cup\{a\in\mathcal L_n:\exists c\in\mathcal C_n^{(s)},\ a<c,\ \omega(Q_a,Q_c)=1\}.
\label{eq:closure-update}
\end{equation}
The fixed point $\mathcal C_n=\mathcal C_n^{(s^\star)}$ is the \emph{measurement-relevant closure} of $M_n$.
Rotations in $\mathcal C_n$ are materialized before $M_n$ in their original circuit order, while the remaining set $\mathcal D_n=\mathcal L_n\setminus\mathcal C_n$ remains pending for subsequent boundaries.
The procedure therefore assigns each retained rotation to the first measurement boundary at which it becomes required under the closure rule.

\noindent\textbf{Preservation of Measurement Statistics.}
Let $W_{\mathcal X}$ denote the product of effective rotations in $\mathcal X$ in circuit order.
By construction, any deferred rotation preceding a selected one commutes with it; otherwise, Eq.~\eqref{eq:closure-update} would include it in $\mathcal C_n$, so the pending sequence can be reordered using only commuting swaps as $W_{\mathcal L_n}=W_{\mathcal D_n}W_{\mathcal C_n}$.
Every rotation in $\mathcal D_n$ also commutes with $M_n$, since any directly anticommuting rotation belongs to $\mathcal C_n^{(0)}$.
For either virtual outcome $\tilde b$,
\begin{equation}
\Pi_{\tilde b}(M_n)W_{\mathcal L_n}|\phi\rangle
=W_{\mathcal D_n}\Pi_{\tilde b}(M_n)W_{\mathcal C_n}|\phi\rangle.
\label{eq:closure-correctness}
\end{equation}
Because $W_{\mathcal D_n}$ is unitary, both sides of Eq.~\eqref{eq:closure-correctness} have the same squared norm.
Materializing only $W_{\mathcal C_n}$ therefore produces exactly the same conditional Born probabilities at $M_n$ as immediate materialization of all pending rotations.
After projection, $W_{\mathcal D_n}$ remains the pending operator relating the MASQ state to the fully materialized state, preserving the same invariant at subsequent boundaries.

\noindent\textbf{Stabilizer-Assisted Elimination.}
Some rotations may remain unnecessary at every measurement boundary, and MASQ identifies them using stabilizer-equivalent representations that allow their effects to propagate beyond future measurements.
For $S|\psi\rangle=\epsilon|\psi\rangle$ with $\epsilon\in\{+1,-1\}$ and $\omega(S,Q)=0$, defining $Q'=QS$ gives $Q'|\psi\rangle=\epsilon Q|\psi\rangle$ and, using Eq.~\eqref{eq:pauli-rotation}, $R_{Q'}(\epsilon\vartheta)|\psi\rangle=R_Q(\vartheta)|\psi\rangle$~\cite{Schmitz:arXiv2023}.

For a subsequent rotation or measurement with Pauli axis $O$, the equivalent axis satisfies $\omega(Q',O)=\omega(Q,O)+\omega(S,O)\pmod 2$.
Thus, if $Q$ anticommutes with $O$, a currently valid stabilizer $S$ satisfying $\omega(Q,S)=0$ and $\omega(S,O)=1$ makes $Q'=QS$ commute with $O$, allowing the equivalent rotation to propagate beyond that operation.
MASQ applies this test sequentially through the remaining circuit while updating the valid stabilizer relations as the state evolves.
If the candidate propagates beyond all remaining measurements, it is removed as \emph{measurement-invisible}; otherwise, it is retained for measurement-relevant closure.
Shot-dependent noise and measurement outcomes alter only the eigenvalue signs of these relations, whereas elimination depends only on commutation parity and preserves the subsequent measurement-outcome distribution, although the unmeasured final state may differ.

Stabilizer-equivalent axes are used only to certify elimination and do not replace the original axes of rotations retained in the runtime schedule, avoiding reliance on state-dependent equivalences that may no longer hold after projection.

Together, closure-based deferral and stabilizer-assisted elimination preserve the conditional virtual-outcome distribution at every measurement boundary.
The corresponding physical measurement record is obtained as
\begin{equation}
b_n=\tilde b_n\oplus\omega(M_n,P_{F,n}),
\label{eq:physical-outcome}
\end{equation}
where $P_{F,n}$ is the Pauli frame immediately before $M_n$.
MASQ therefore preserves $P(b_n\mid b_1,\ldots,b_{n-1})$ for every physical-noise realization and measurement prefix.
Thus, by the chain rule, $P(\mathbf b)=\prod_{n=1}^{m}P(b_n\mid b_1,\ldots,b_{n-1})$ is identical to that of baseline exact simulation, preserving the complete measurement-record distribution.

With $r_{\max}$ denoting MASQ's peak active width, reducing the peak from $k_{\max}$ to $r_{\max}$ decreases the peak dense-state size by a factor of $2^{k_{\max}-r_{\max}}$.

\begin{algorithm}[t]
\caption{MASQ compilation and sampling}
\label{alg:shot_sampling}
\footnotesize 
\setlength{\baselineskip}{1.1em} 
\begin{algorithmic}

\Function{\textsc{Compile\_Schedule}}{$\mathcal Q$}
\Repeat
\State $\mathcal I\gets\emptyset$, Update the stabilizer tables $\mathcal T_a$ for the current $\mathcal Q$
\For{\textbf{each} non-Clifford rotation $R_{Q_a}$ indexed by $a$ in $\mathcal Q$}
\State $P\gets Q_a$; $\mathcal T\gets\mathcal T_a$; $\texttt{removable}\gets\textbf{true}$
\For{\textbf{each} subsequent operation $o$}
    \State \textbf{if} $o$ is a Clifford $C$ \textbf{then} $P\gets CPC^\dagger$
    \State \textbf{else if} $o$ has Pauli axis $O$ and $\omega(P,O)=1$ \textbf{then}
    \State \hspace{\algorithmicindent}\textbf{if} a valid $S\in\langle\mathcal T\rangle$ exists \textbf{then} $P\gets PS$
    \State \hspace{\algorithmicindent}\textbf{else} $\texttt{removable}\gets\textbf{false}$; \textbf{break}
    \State {Update $\mathcal T$ with $o$ to track valid Pauli relations}
\EndFor
\State \textbf{if} $\texttt{removable}$ \textbf{then} $\mathcal I\gets\mathcal I\cup\{a\}$
\EndFor
\State $\mathcal Q\gets\mathcal Q\setminus\{R_{Q_a}\mid a\in\mathcal I\}$
\Until{$\mathcal I=\emptyset$}

\State Split $\mathcal Q$ into $\{(G_n,M_n)\}_{n=1}^{m}$; $\mathcal L\gets[\,]$
\For{$n=1$ \To $m$}
\State Append rotation indices of $G_n$ to $\mathcal L$
\State $\mathcal C_n\gets\{a\in\mathcal L\mid\omega(Q_a,M_n)=1\}$
\Repeat
\State $\mathcal C_n\gets\mathcal C_n\cup\{a\in\mathcal L\mid\exists c\in\mathcal C_n,\ a<c,\ \omega(Q_a,Q_c)=1\}$
\Until{$\mathcal C_n$ is unchanged}
\State Order $\mathcal C_n$ by circuit index; $\mathcal L\gets\mathcal L\setminus\mathcal C_n$
\EndFor
\State \Return $\texttt{Rotation\_Plan}\gets\bigl(\{G_n,M_n,\mathcal C_n\}_{n=1}^{m},\{(a,Q_a,\theta_a)\}\bigr)$
\EndFunction

\vspace{0.2em} 

\Function{\textsc{Sample\_Shots}}{$\texttt{Rotation\_Plan},\mathcal J$}
\For{\textbf{each} $i\in\mathcal J$}
\State $F\gets I$; initialize $|\phi\rangle$ and the pending rotation-sign context
\For{$n=1$ \To $m$}
\For{\textbf{each} operation $o$ in $G_n$}
\State \textbf{if} $o$ is a Pauli operation $E$ \textbf{then} $F\gets EF$
\State \textbf{if} $o$ is rotation $a$ \textbf{then} Track $\vartheta_a$ and defer it
\EndFor
\For{\textbf{each} $a\in\mathcal C_n$ in circuit order}
\State Apply $R_{Q_a}(\vartheta_a)$ to $|\phi\rangle$
\EndFor
\State Compute $(p_0,p_1)$ for $M_n$ and draw virtual outcome $\tilde b_n^{(i)}$
\State $b_n^{(i)}\gets\tilde b_n^{(i)}\oplus\omega(M_n,F)$; project $|\phi\rangle$ onto $\tilde b_n^{(i)}$ and normalize
\State Demote measured coordinates when applicable
\EndFor
\EndFor
\State \Return $\{\mathbf b^{(i)}\}_{i\in\mathcal J}$
\EndFunction

\end{algorithmic}
\end{algorithm}

\vspace{-0.2em}
\subsection{Algorithm Implementation}\label{sec:masq-implementation}
\vspace{-0.3em}
MASQ consists of compile-time rotation scheduling, schedule-driven shot sampling, and adaptive boundary caching.
Algorithm~\ref{alg:shot_sampling} removes measurement-invisible rotations, determines the measurement boundary of each retained rotation, and generates a shared \texttt{Rotation\_Plan}.
The same plan is reused across all shots, while shot-dependent Pauli-frame effects are handled during runtime sampling.

\noindent\textbf{Compile-Time Rotation Scheduling.}
The first phase of \textsc{Compile\_Schedule} performs stabilizer-assisted elimination.
The compiler maintains independent Pauli eigenstate relations in a tableau $\mathcal T$ and tests each candidate rotation using the relations valid at its circuit position.
When the transported candidate axis $P$ anticommutes with a subsequent Pauli axis $O$, the compiler searches for a valid stabilizer $S\in\langle\mathcal T\rangle$ satisfying $\omega(P,S)=0$ and $\omega(S,O)=1$.
If no such relation exists, the rotation is retained. Because removing measurement-invisible rotations can restore discarded Pauli relations, the stabilizer tables are recomputed after each pass, and the procedure repeats until convergence.

The second phase constructs the measurement-relevant closures of the remaining rotations.
Indices without an assigned boundary are maintained in the pending list $\mathcal L$.
After the rotations of segment $G_n$ are appended to $\mathcal L$, the compiler computes $\mathcal C_n$ according to Eq.~\eqref{eq:closure-update}, assigns its rotations to $M_n$ in circuit order, and carries the remaining indices to the next boundary.
The resulting \texttt{Rotation\_Plan} stores $\{G_n,M_n,\mathcal C_n\}_{n=1}^{m}$ together with the axes, angles, and circuit identities of the retained rotations.

\noindent\textbf{Schedule-Driven Shot Sampling.}
Each shot begins with the identity Pauli frame and the initial frame-factored state.
Within each segment $G_n$, shot-dependent Pauli errors and feed-forward corrections are accumulated in $F$, while non-Clifford rotations remain pending.
MASQ fixes each effective angle $\vartheta_a$ at the rotation's original circuit position using Eq.~\eqref{eq:rotation-frame} and preserves it until materialization.
Immediately before $M_n$, only the rotations assigned to $\mathcal C_n$ are materialized in their original circuit order.
MASQ then computes the virtual Born probabilities $(p_0,p_1)$ and samples $\tilde b_n$.
The recorded physical bit is obtained from Eq.~\eqref{eq:physical-outcome}, after which the active state is projected onto $\tilde b_n$ and eligible measured coordinates are returned to the dormant representation.
Repeating this procedure over all boundaries generates the measurement record $\mathbf b^{(i)}$.

\vspace{-0.2em}
\subsection{Adaptive Boundary Caching}\label{sec:adaptive-caching}
\vspace{-0.2em}

By restricting each boundary to the non-Clifford information required for its Born probabilities, MASQ defines compact probability-determining contexts that can recur across shots.
Adaptive caching reuses the corresponding conditional Born probabilities when such contexts recur.

For shot $i$, $\Gamma_n^{(i)}$ represents the effective angle sign history of rotations materialized through boundary $n$, with signs fixed at their original positions.
MASQ constructs the cache key as
\begin{equation}
\texttt{Key}_n^{(i)}=\left(n,\,\Gamma_n^{(i)},\,\tilde b_1^{(i)},\ldots,\tilde b_{n-1}^{(i)}\right).
\label{eq:boundary-key}
\end{equation}
For a fixed \texttt{Rotation\_Plan}, identical keys represent the same sequence of materialized rotations and prior virtual projections, yielding the same conditional Born probabilities at $M_n$. The \texttt{Cache} maps each key to $(p_0,p_1)$.

On a cache hit, MASQ skips active-state reconstruction, non-Clifford rotation application, and Born probability evaluation and directly samples $\tilde b_n$ from the cached probabilities.
The physical outcome $b_n$ is then obtained from the current Pauli frame using Eq.~\eqref{eq:physical-outcome}.
On a cache miss, MASQ evaluates the shot through the uncached sampling path, adds the newly computed boundary probabilities to the \texttt{Cache}, and completes the measurement record.
At capacity, further cache growth cannot reduce misses, so MASQ switches to uncached sampling when Eq.~\eqref{eq:cache_now} is violated.
Because caching only reuses an already determined conditional probability distribution and each outcome is sampled independently from that distribution, caching does not alter the shot distribution.

\begin{algorithm}[t]
\vskip -1pt
\caption{Adaptive boundary caching}
\label{alg:cached-sampling}
\footnotesize
\setlength{\baselineskip}{1.1em}
\begin{algorithmic}

\Function{\textsc{Adaptive\_Sample\_Shots}}{$\texttt{Rotation\_Plan},N_{\mathrm{shot}}$}
\State Run $N_{\mathrm{probe}}$ uncached probe shots; measure $\tau_{\mathrm{base}}$ and initialize \texttt{Cache}
\For{$i=N_{\mathrm{probe}}+1$ \To $N_{\mathrm{shot}}$}
\State $F\gets I$; initialize sign context $\Gamma$ and virtual-outcome history
\For{$n=1$ \To $m$}
\State Process $G_n$ to update $F$ and $\Gamma$; construct $\texttt{Key}_n^{(i)}$ using Eq.~\eqref{eq:boundary-key}
\If{$\texttt{Key}_n^{(i)}\in\texttt{Cache}$}
\State Draw and append $\tilde b_n^{(i)}$ from $\texttt{Cache}[\texttt{Key}_n^{(i)}]$
\State $b_n^{(i)}\gets\tilde b_n^{(i)}\oplus\omega(M_n,F)$
\Else
\State Reconstruct and complete shot $i$ using the sampled prefix
\State \textbf{if} $\operatorname{size}(\texttt{Cache}) < \operatorname{capacity}(\texttt{Cache})$ \textbf{ then} update \texttt{Cache}
\State \textbf{break}
\EndIf
\EndFor
\State Update $\mu$, $\tau_{\mathrm{hit}}$, $\tau_{\mathrm{miss}}$, and $D(i)$
\If{$(i-N_{\mathrm{probe}})\mathbin{\%} N_{\mathrm{check}}=0$ and $\mu\tau_{\mathrm{miss}}+(1-\mu)\tau_{\mathrm{hit}}>\tau_{\mathrm{base}}$}
\State {\textbf{if} \texttt{Cache} is full \textbf{ then} run remaining shots uncached; \textbf{break}}
\State Fit $D(j)\approx\lambda j^\beta$; $\widehat N_{\mathrm{miss}}\gets\lambda(N_{\mathrm{shot}}^\beta-i^\beta)$
\If{$\widehat N_{\mathrm{miss}}(\tau_{\mathrm{miss}}-\tau_{\mathrm{hit}})>(N_{\mathrm{shot}}-i)(\tau_{\mathrm{base}}-\tau_{\mathrm{hit}})$}
\State Run the remaining shots using uncached \textsc{Sample\_Shots} \textbf{break}
\EndIf
\EndIf
\EndFor
\State \Return $\mathbf b^{(1)},\ldots,\mathbf b^{(N_{\mathrm{shot}})}$
\EndFunction

\end{algorithmic}
\vspace{-0.2em}
\end{algorithm}

Algorithm~\ref{alg:cached-sampling} summarizes adaptive caching, which begins with $N_{\mathrm{probe}}$ uncached probe shots to measure $\tau_{\mathrm{base}}$ and initialize the \texttt{Cache}; these shots are retained in the final measurement records.
During subsequent execution, $\tau_{\mathrm{hit}}$ and $\tau_{\mathrm{miss}}$ denote the mean times of all-hit and miss-containing shots, respectively, while $\mu$ denotes the observed miss ratio.
Caching is currently beneficial when
\begin{equation}
\mu\tau_{\mathrm{miss}}
+
(1-\mu)\tau_{\mathrm{hit}}
\leq
\tau_{\mathrm{base}}.
\label{eq:cache_now}
\end{equation}

Before capacity is reached, MASQ does not immediately disable caching when Eq.~\eqref{eq:cache_now} is violated because future misses may decrease as the \texttt{Cache} grows.
Let $D(j)$ denote the number of distinct boundary keys observed after $j$ shots.
MASQ fits $D(j)\approx\lambda j^\beta$ for $1\leq j\leq i$ using least squares in the log domain and estimates the remaining miss-containing shots as $\widehat N_{\mathrm{miss}}=\lambda(N_{\mathrm{shot}}^\beta-i^\beta)$.
Caching is disabled when
\begin{equation}
\widehat N_{\mathrm{miss}}(\tau_{\mathrm{miss}}-\tau_{\mathrm{hit}})
>
(N_{\mathrm{shot}}-i)(\tau_{\mathrm{base}}-\tau_{\mathrm{hit}}),
\label{eq:cache_switch}
\end{equation}
where the two sides estimate the future miss overhead and the maximum remaining cache-hit saving, respectively.
Once disabled, remaining shots use uncached sampling.

\noindent\textbf{Complexity.}
The compile-time cost of MASQ is dominated by stabilizer-assisted elimination and measurement-relevant closure construction.
For $N$ virtual qubits, $R$ non-Clifford rotations, $m$ measurements, and $L$ circuit operations, these require $O(R^2LN^2)$ and $O(mR^2)$ time in the worst case, respectively.
The resulting compile-time complexity is $O(R^2LN^2+mR^2)$ with $O(N^2+R^2+L+m)$ space, and this one-time cost is shared across all shots.
Uncached sampling requires $O(L+(R+m)2^{r_{\max}})$ time per shot and $O(2^{r_{\max}})$ dense-state memory.
Caching avoids dense-state operations on a hit but does not increase the worst-case runtime of $O(N_{\mathrm{shot}}[L+(R+m)2^{r_{\max}}])$; with at most $D_{\max}$ cached contexts and $S_{\mathrm{key}}$ storage per key, the total runtime space is $O(2^{r_{\max}}+N+L+R+m+D_{\max}S_{\mathrm{key}})$.

\section{Experimental Evaluation}\label{sec:exp}

\subsection{Experimental Setup}\label{sec:exp_setup}
\vspace{-0.2em}

We evaluate MASQ against the exact near-Clifford samplers Clifft~\cite{Chase:arXiv2026}, Tsim~\cite{Haenel:arXiv2026}, and SymFT~\cite{Fang:arXiv2026}, with Stim~\cite{Gidney:Quantum2021} additionally evaluated on the pure-Clifford benchmark.
Clifft is evaluated with its default HIR optimization pipeline, including active-width squeezing.
The suite comprises 19 QEC circuits based on Clifft's benchmarks: surface-code memories with coherent $R_Z$ errors~\cite{Tuloup:arXiv2026} ($d=3,5,7,9$) or $R_X$ errors ($d=3$), magic-state distillation~\cite{Sales:Nature2025}, magic-state cultivation~\cite{Gidney:arXiv2024} ($d=3,5$), and a pure-Clifford surface-code memory ($d=7$).
We extend the original suite with multi-round $R_Z$ circuits at $d=7$ and a $d=9$ circuit to evaluate larger instances.
Coherent variants use $R_{Z/X}(0.02)$ at the corresponding noise locations.

CPU experiments use 16 physical cores of an Intel Xeon Gold 6246R with 614\,GiB host memory; Tsim and SymFT are additionally evaluated on an RTX A6000.
MASQ uses $N_{\mathrm{probe}}=512$ shots, 64 shots per word for batched execution, and $N_{\mathrm{check}}=2048$ shots throughout.
Every simulator is subject to a 600\,s compilation limit per circuit, and sampling throughput and compilation time are reported separately.
For runs reaching saturation, capacity, or halt after $N_{\mathrm{warm}}$ shots, throughput is measured over the next $N_{\mathrm{warm}}$ shots.
The \texttt{Cache} is implemented as a per-core hash table capped at 2\,GiB to bound memory overhead in multi-core execution. Thus, the reported points are averaged across cores.

\subsection{Experimental Results}\label{sec:exp_results}

\begin{table*}[!t]
\centering
\caption{Shot-sampling throughput (shots/s) and MASQ characteristics across 19 QEC circuits at a Pauli error rate of $10^{-3}$.
$d$ and $r$ denote code distance and measurement rounds; 
$k_{\max}$ and $r_{\max}$ denote the peak active widths for Clifft after HIR reordering and MASQ, respectively;
$t_{\mathrm{compile}}$ is the MASQ compilation time (s).
Accuracy reports maximum absolute errors in Born probability and log probability, $|\Delta P|$ and $|\Delta\log P|$, against the baseline.
Parentheses report caching status or saturation/capacity point, when applicable.}
\label{tab:cpu}
\vskip -2pt
\renewcommand{\arraystretch}{1.2}
\setlength{\tabcolsep}{3pt}
\resizebox{\textwidth}{!}{%
\begin{tabular}{c|c|c|c|c|c|c|c|c|c|c|c|c|c|c|c|c}
\Xhline{1pt}
\rowcolor[HTML]{F2F2F2}
\multicolumn{5}{c|}{\textbf{Benchmark}} &
\multicolumn{2}{c|}{\textbf{GPU Sampling Throughput}} &
\multicolumn{5}{c|}{\textbf{CPU Sampling Throughput}} &
\multicolumn{1}{c|}{~} &
\multicolumn{4}{c}{\textbf{MASQ Characteristics}} \\
\Xhline{0.6pt}
\rowcolor[HTML]{F2F2F2}
\textbf{Circuit} & \textbf{Variant} & \textbf{$d$} &
\textbf{$r$} & \textbf{Qubits} &
\textbf{Tsim}~\cite{Haenel:arXiv2026} &
\textbf{SymFT}~\cite{Fang:arXiv2026} &
\textbf{Clifft}~\cite{Chase:arXiv2026} &
\textbf{Tsim}~\cite{Haenel:arXiv2026} &
\textbf{SymFT}~\cite{Fang:arXiv2026} &
\textbf{Stim}~\cite{Gidney:Quantum2021} &
\textbf{MASQ (Caching)} &
$\mathbf{k_{\max}}$ & $\mathbf{r_{\max}}$ &
$\mathbf{t_{\mathrm{compile}}}$ &
\textbf{Max. $|\Delta P|$} &
\textbf{Max. $|\Delta\log P|$} \\
\Xhline{0.8pt}

\multirow[c]{13}{*}{Surface} &
\multirow[c]{13}{*}{$R_Z$ error}
& 9 & 1 & 188
& $1.59{\times}10^{7}$ & OOM$^{*}$ & OOM
& $\mathbf{2.66{\times}10^{8}}$ & OOM & --
& $1.16{\times}10^{8}$ (Halted) & 40 & 0 & 17.99
& -- & -- \\

\Xcline{3-17}{0.4pt}
& & \multirow[c]{4}{*}{7} & 4 & \multirow[c]{4}{*}{118}
& DNC & OOM & OOM & DNC & OOM & --
& $\mathbf{3.02{\times}10^{-1}}$ (Halted) & 45 & 25 & 32.84
& -- & -- \\

& & & 3 &
& DNC & OOM & OOM & DNC & OOM & --
& $\mathbf{5.52{\times}10^{-1}}$ (Halted) & 45 & 25 & 24.72
& -- & -- \\

& & & 2 &
& DNC & OOM & OOM & DNC & OOM & --
& $\mathbf{3.07}$ (Halted) & 45 & 25 & 16.39
& -- & -- \\

& & & 1 &
& $2.75{\times}10^{7}$ & $2.58{\times}10^{-1}$ & $1.73$
& $\mathbf{4.65{\times}10^{8}}$ & $5.53$ & --
& $1.96{\times}10^{8}$ (Halted) & 24 & 0 & 3.90
& $0.00{\rm e}{+00}$ & $0.00{\rm e}{+00}$ \\

\Xcline{3-17}{0.4pt}
& & \multirow[c]{4}{*}{5} & 7 & \multirow[c]{4}{*}{64}
& DNC & $3.19$ & $3.08$ & DNC & $1.95{\times}10^{1}$ & --
& $\mathbf{1.98{\times}10^{4}}$ (Halted) & 22 & 13 & 9.72
& $3.73{\times}10^{-25}$ & $2.06{\times}10^{-13}$ \\

& & & 5 &
& DNC & $5.13$ & $4.58$ & DNC & $2.94{\times}10^{1}$ & --
& $\mathbf{3.14{\times}10^{4}}$ (Halted) & 22 & 13 & 6.97
& $9.10{\times}10^{-24}$ & $2.49{\times}10^{-13}$ \\

& & & 3 &
& DNC & $1.05{\times}10^{1}$ & $9.20$
& DNC & $5.88{\times}10^{1}$ & --
& $\mathbf{7.45{\times}10^{4}}$ (Halted) & 22 & 13 & 4.22
& $9.31{\times}10^{-23}$ & $7.46{\times}10^{-14}$ \\

& & & 1 &
& $9.90{\times}10^{7}$ & $9.17{\times}10^{4}$
& $1.61{\times}10^{5}$ & $\mathbf{1.01{\times}10^{9}}$
& $3.06{\times}10^{5}$ & --
& $2.72{\times}10^{8}$ (Halted) & 12 & 0 & 0.67
& $0.00{\rm e}{+00}$ & $0.00{\rm e}{+00}$ \\

\Xcline{3-17}{0.4pt}
& & \multirow[c]{4}{*}{3} & 7 & \multirow[c]{4}{*}{26}
& DNC & $1.52{\times}10^{6}$ & $1.33{\times}10^{6}$
& DNC & $2.53{\times}10^{6}$ & --
& $\mathbf{4.88{\times}10^{6}}$ ($2.24{\times}10^{8}$) & 7 & 5 & 0.79
& $1.95{\times}10^{-18}$ & $2.13{\times}10^{-14}$ \\

& & & 5 &
& DNC & $2.04{\times}10^{6}$ & $1.93{\times}10^{6}$
& DNC & $3.24{\times}10^{6}$ & --
& $\mathbf{1.18{\times}10^{7}}$ ($1.46{\times}10^{7}$)
& 7 & 5 & 0.56
& $3.25{\times}10^{-18}$ & $1.07{\times}10^{-14}$ \\

& & & 3 &
& DNC & $2.92{\times}10^{6}$ & $3.19{\times}10^{6}$
& DNC & $5.85{\times}10^{6}$ & --
& $\mathbf{3.10{\times}10^{7}}$ ($7.01{\times}10^{6}$)
& 7 & 5 & 0.35
& $3.47{\times}10^{-18}$ & $7.11{\times}10^{-15}$ \\

& & & 1 &
& $2.27{\times}10^{8}$ & $5.32{\times}10^{6}$
& $1.95{\times}10^{7}$ & $\mathbf{3.33{\times}10^{9}}$
& $3.50{\times}10^{7}$ & --
& $8.00{\times}10^{8}$ (Halted) & 4 & 0 & 0.08
& $0.00{\rm e}{+00}$ & $0.00{\rm e}{+00}$ \\

\Xhline{0.6pt}
\multirow[c]{2}{*}{Surface} &
\multirow[c]{2}{*}{$R_X$ error} &
\multirow[c]{2}{*}{3}
& 3 & \multirow[c]{2}{*}{26}
& DNC & $2.92{\times}10^{4}$ & $6.46{\times}10^{4}$
& DNC & $1.75{\times}10^{4}$ & --
& $\mathbf{5.10{\times}10^{6}}$ ($6.09{\times}10^{6}$) & 14 & 12 & 0.38
& $9.11{\times}10^{-18}$ & $1.07{\times}10^{-14}$ \\

& & & 1 &
& DNC & $8.85{\times}10^{4}$ & $1.90{\times}10^{5}$
& DNC & $6.02{\times}10^{4}$ & --
& $\mathbf{2.31{\times}10^{7}}$ ($9.79{\times}10^{6}$)
& 14 & 11 & 0.10
& $1.17{\times}10^{-17}$ & $7.11{\times}10^{-15}$ \\

\Xhline{0.6pt}
\multirow[c]{2}{*}{Cultivation} &
\multirow[c]{2}{*}{--}
& 5 & -- & 42
& DNC & $2.21{\times}10^{5}$ & ${6.85{\times}10^{5}}$
& DNC & $5.24{\times}10^{5}$ & --
& $\mathbf{1.83{\times}10^{6}}$ ($1.60{\times}10^{7}$)${}^\dagger$ & 10 & 10 & 0.14
& $1.55{\times}10^{-25}$ & $2.57{\times}10^{-14}$ \\

& & 3 & -- & 15
& $1.89{\times}10^{4}$ & $4.39{\times}10^{6}$
& $9.44{\times}10^{6}$ & $1.37{\times}10^{2}$
& $2.93{\times}10^{7}$ & --
& $\mathbf{4.57{\times}10^{7}}$ ($1.83{\times}10^{6}$)
& 4 & 4 & 0.02
& $1.76{\times}10^{-18}$ & $6.66{\times}10^{-16}$ \\

\Xhline{0.6pt}
Distillation & -- & 5 & -- & 85
& $6.85{\times}10^{5}$ & $2.47{\times}10^{6}$
& $5.19{\times}10^{6}$ & $5.71{\times}10^{4}$
& $1.08{\times}10^{7}$ & --
& $\mathbf{3.97{\times}10^{7}}$ ($2.10{\times}10^{5}$)
& 5 & 4 & 0.04
& $4.29{\times}10^{-28}$ & $7.11{\times}10^{-15}$ \\

\Xhline{0.6pt}
Surface & Clifford & 7 & 7 & 118
& $1.30{\times}10^{6}$ & $5.95{\times}10^{5}$
& $2.60{\times}10^{6}$ & $1.40{\times}10^{7}$
& $5.75{\times}10^{6}$ & $1.52{\times}10^{7}$
& $\mathbf{3.52{\times}10^{7}}$ (Halted) & 0 & 0 & 0.09
& $0.00{\rm e}{+00}$ & $0.00{\rm e}{+00}$ \\

\Xhline{0.5pt}
\end{tabular}%
}

{\scriptsize\raggedleft
$^{*}$\textit{Note:} OOM = out of memory; DNC = compilation $>600$\,s; -- = unsupported or not evaluated.
$^{\dagger}$2\,GiB per-core capacity reached.\par}
\vskip -14pt
\end{table*}

\subsubsection{Shot-Sampling Throughput}
Table~\ref{tab:cpu} compares all CPU and GPU configurations at a Pauli error rate of {$10^{-3}$}.
MASQ samples all 19 circuits and achieves the highest throughput on 15, whereas Clifft and SymFT each encounter OOM on four circuits and Tsim exceeds the compilation limit on 12.
Of these 12 DNC cases, 11 are coherent-error surface codes, consistent with Tsim's exponential dependence on the number of non-Clifford gates~\cite{Haenel:arXiv2026}.
For surface $R_Z$ at $d=5,r=3$, MASQ reaches {$7.45{\times}10^4$}\,shots/s, outperforming Clifft and CPU SymFT by {$8{,}098\times$} and {$1{,}267\times$}, respectively.
For distillation, MASQ reaches {$3.97{\times}10^7$}\,shots/s, corresponding to {$7.65\times$} and {$3.68\times$} speedups.
The gains extend to coherent $R_X$ errors: at $r=1$, MASQ achieves {$2.31{\times}10^7$}\,shots/s, {$122\times$} faster than Clifft and {$384\times$} faster than CPU SymFT.
For pure Clifford, MASQ reaches {$3.52{\times}10^7$}\,shots/s, {$2.32\times$} over Stim.

\begin{figure}[t]
\vskip 4pt
\centering
\includegraphics[width=\columnwidth]{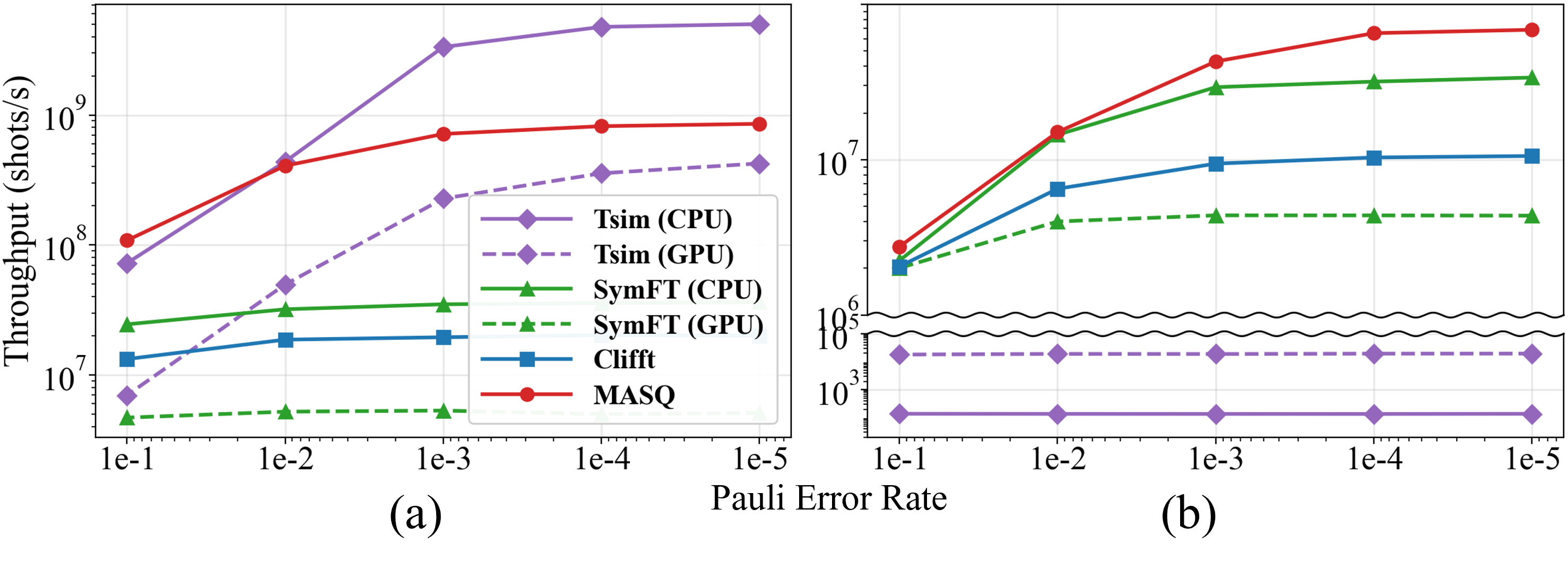}
\vskip -6pt
\caption{Throughput versus Pauli error rate for (a) surface $R_Z$ ($d=3,r=1$) and (b) cultivation ($d=3$); MASQ caching is halted in (a) and active in (b).}
\label{fig:p_sweep}
\end{figure}

CPU Tsim remains fastest on all single-round $R_Z$ circuits, outperforming MASQ by {$2.29$--$4.16\times$}.
For the executable $d=3,5,7$ cases, Tsim exceeds Clifft by $171\times$ to $2.69{\times}10^8\times$, whereas its advantage over MASQ remains only {$2.37$--$4.16\times$}.
At $d=9$, MASQ also completes sampling within {$2.29\times$} of Tsim despite Clifft and SymFT encountering OOM.
These MASQ runs have $r_{\max}=0$, making the remaining gap consistent with residual frame and per-shot control overhead once dense-state processing is eliminated.
\refFigure{fig:p_sweep} examines throughput across the Pauli error rate for two representative $d=3$ cases where Tsim successfully compiles.
In Fig.~\ref{fig:p_sweep}(a), CPU Tsim leads at low error rates for the single-round surface $R_Z$ circuit ($d=3,r=1$), but the curves cross at $p=0.011$ and approximately $3.9{\times}10^8$\,shots/s as Tsim throughput decreases more sharply, with MASQ leading at higher error rates even within this Tsim-compilable subset.
In Fig.~\ref{fig:p_sweep}(b), for $d=3$ cultivation with caching enabled, MASQ maintains the highest throughput throughout the sweep, with its advantage widening at lower error rates, consistent with greater reuse of cached measurement contexts.

\subsubsection{Active-State Reduction}

\begin{figure}[t]
\vskip 4pt
\centering
\includegraphics[width=\columnwidth]{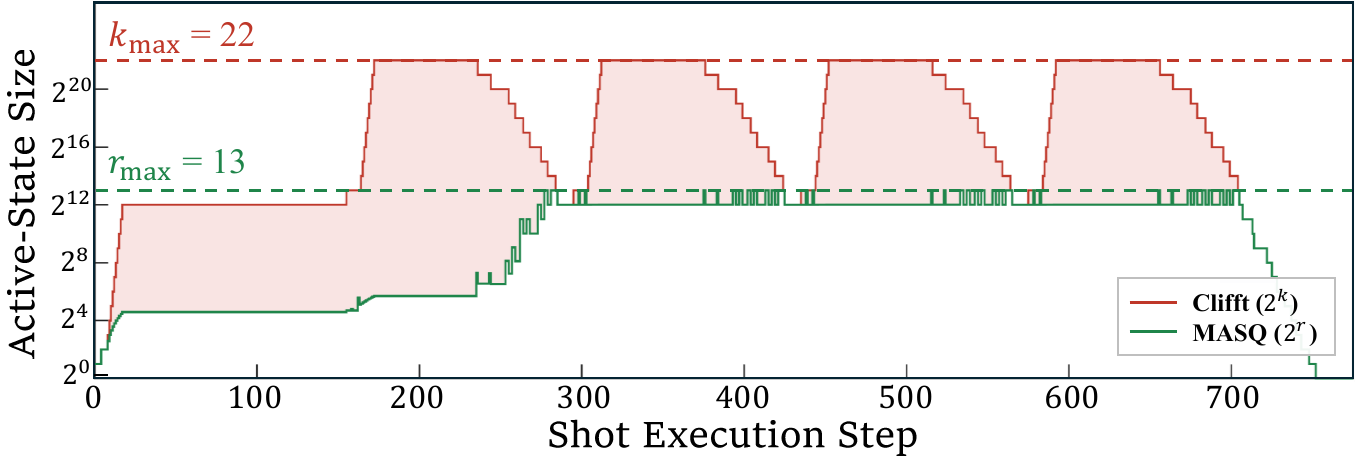}
\caption{Active-state size throughout the surface-code memory circuit with coherent $R_Z$ errors ($d=5,r=5$).
MASQ reduces the peak amplitude count from $2^{22}$ to {$2^{13}$}.}
\label{fig:active}
\end{figure}

MASQ reduces the peak active width in most near-Clifford benchmarks, with particularly large reductions on multi-round $R_Z$ circuits: {$22\rightarrow13$} at $d=5$ and {$45\rightarrow25$} at $d=7$.
Fig.~\ref{fig:active} illustrates the $d=5,r=5$ case, where Clifft repeatedly reaches $2^{22}$ amplitudes while MASQ limits the peak to {$2^{13}$}, reducing the peak dense-state size by {$512\times$}.
MASQ's one-time compilation takes at most 32.84~s across the evaluated circuits, and the resulting schedule is shared across all shots to amortize this cost.

\subsubsection{Adaptive Caching Performance}

Table~\ref{tab:cpu} shows that caching remains enabled for the $d=3$ multi-round $R_Z$ circuits, both $R_X$ variants, and the magic-state circuits, but is disabled elsewhere.
Reported saturation/capacity points range from {$2.10{\times}10^5$ to $2.24{\times}10^8$} shots, motivating runtime cache control rather than a uniform caching policy.
Notably, $d=3$ cultivation has $k_{\max}=r_{\max}=4$, so its throughput improvement is achieved without reducing the peak active width.

\noindent\textbf{Correctness Validation.}
\refTable{tab:cpu} reports the maximum absolute and log Born-probability errors, $|\Delta P|$ and $|\Delta\log P|$, between MASQ and Clifft over all sampled shots.
Comparisons match noise realizations and measurement outcomes, including cached execution where applicable.
Across 15 circuits, the maximum $|\Delta P|$ is {$1.17{\times}10^{-17}$} for $R_X$ {($d=3,r=1$)}, within double-precision roundoff, where {$|\Delta\log P|=7.11{\times}10^{-15}$}.

\vspace {-0.2em}
\section{Conclusion}
\vspace {-0.3em}
This paper presented MASQ, an exact measurement-adaptive sampler that reduces the active-state bottleneck in near-Clifford QEC simulation through selective materialization of non-Clifford rotations.
Measurement-relevant closures defer rotations unnecessary for the current measurement, while stabilizer-assisted elimination removes rotations invisible to all future measurements; adaptive caching further reduces repeated work across shots.
Across 19 QEC benchmarks, MASQ reduces peak active width by up to 40 active virtual qubits and achieves up to $8{,}098\times$ CPU throughput speedup over Clifft on multi-round circuits, while completing circuits for which existing exact simulators hit memory or compilation limits.
These results demonstrate that measurement-adaptive control of non-stabilizer information can substantially improve the practical scalability of exact QEC sampling while preserving the measurement-record distribution exactly.

\newpage
\bibliographystyle{IEEEtran}
\bibliography{reference}

\end{document}